\documentclass[prl,twocolumn,showpacs]{revtex4}
\usepackage{graphicx}
\usepackage{color}
\usepackage{bm}
\usepackage{longtable}
\usepackage{graphics}
\usepackage{amsmath}
\usepackage{bm}
\usepackage{amsfonts}
\usepackage{amssymb}
\usepackage[T1]{fontenc}
\usepackage[latin9]{inputenc}
\usepackage{esint}
\begin{document}

\title{A New Method of Classification of Pure Tripartite Quantum States}

\author{Jyoti Faujdar}
\thanks{faujdar.jyoti@gmail.com}
\affiliation{Indian Institute of Technology Jodhpur, Jodhpur
-342011, India}
\author{Anoopa Joshi}
\thanks{anoopa.mymails@gmail.com}
\affiliation{Indian Institute of Technology Jodhpur, Jodhpur
-342011, India}
\author{Satyabrata Adhikari}
\thanks{tapisatya@gmail.com}
\affiliation{Institute of Physics, Bhubaneswar-751005, India}

\date{\today}

\begin{abstract}
\noindent The classification of the multipartite entanglement is
an important problem in quantum information theory. We propose a
class of two qubit mixed states $\sigma_{AB}=
p|\chi_{1}\rangle\langle\chi_{1}|\otimes\rho_{1}+(1-p)|\chi_{2}\rangle\langle\chi_{2}|\otimes\rho_{2}$,
where $|\chi_{1}\rangle=\alpha|0\rangle+\beta|1\rangle$,
$|\chi_{2}\rangle=\beta|0\rangle+(-1)^{n}\alpha|1\rangle$. We have
shown that the state $\sigma_{AB}$ represent a classical state
when $n$ is odd while it represent a non-classical state when $n$
is even. The purification of the state $\sigma_{AB}$ is studied
and found that the purification is possible if the spectral
decomposition of the density matrices $\rho_{1}$ and $\rho_{2}$
represent pure states. We have established a relationship between
three tangle, which measures the amount of entanglement in three
qubit system and the quantity $\langle\chi_{1}|\chi_{2}\rangle$,
which identifies whether the two qubit mixed state is classical or
non-classical. The three qubit purified state is then classified
as a separable or biseparable or W-type or GHZ-type state using
the quantum correlation, measured by geometric discord, of its
reduced two qubit density matrix.
\end{abstract}


\maketitle

\section{I. Introduction}
Quantum Entanglement \cite{EPR,bell} is an unavoidable part of
quantum information theory and is used in many quantum information
processing task such as teleportation \cite{teleport}, superdense
coding \cite{dense}, cryptography \cite{review1}, remote state
preparation \cite{pati}. The characterization and quantification
of pure as well as mixed two qubit entanglement are well studied
\cite{review}.

A general notion of entanglement in non-classical two-qubit
systems is proposed by Ollivier and Zurek \cite{oll}. This measure
of quantum correlation in non-classical system captured by quantum
discord. For bipartite system, quantum discord is defined as the
difference between the total mutual information and classical
information \cite{hen,luo}. A generalization for multipartite
systems is facilitated by the formulation of discord as a distance
measure coded by the difference in relative entropy of the quantum
state in question with that of the nearest classical state
\cite{Williamson}. In a similar spirit the notion of normalised
geometric discord has also been introduced \cite{dakic}, which is
defined as
\begin{eqnarray}
D_{G}(\rho_{AB}) &=& 2Min_{\chi_{AB}\in
T}\|\rho_{AB}-\chi_{AB}\|^{2}_{HS} \label{discord}
\end{eqnarray}
where $T$ denote the set of zero discord states and $\|.\|_{HS}$
denote Hilbert Schmidt norm. A common noise processes can be used
to realize discordant states experimentally \cite{lanyon}.

Unlike two qubit state, the structure of three qubit system is not
simple. In three qubit system, there are three kind of states:
separable, biseparable and genuine entangled states. It is very
important to classify these three type of states. Dur et.al.
\cite{dur} have shown that there are six different equivalence
classes of three qubit pure states in terms of stochastic local
operation and classical communication. In addition, they found
that GHZ-class and W-class, where
\begin{eqnarray}
|GHZ\rangle=\lambda|000\rangle + \mu
|111\rangle,\lambda^{2}+\mu^{2}=1\\
|W\rangle=\gamma|100\rangle+\delta|010\rangle+\nu|001\rangle,
\gamma^{2}+\delta^{2}+\nu^{2}=1 \label{ghzw}
\end{eqnarray}
\noindent are two inequivalent class of three qubit entangled
states. W state is robust against loss of one qubit i.e. if one of
the three qubits is lost, the state of the remaining 2-qubit
system is still entangled whereas GHZ state is fully separable
after loss of one qubit. Y. Gao et.al. \cite{gao} have studied the
method of preparation of maximally entangled multipartite GHZ and
W states. For three qubit states, there exist Bell inequalities
that can be violated by W states and not violated by GHZ states
and vice-versa \cite{brunner}.

Recently, Gour and Wallach provided a classification of
multiparticle entanglement in terms of equivalence classes of
states under stochastic local operations and classical
communication (SLOCC) \cite{gour}. A method is proposed to
characterize the genuine multisite entanglement in isotropic
square spin-$\frac{1}{2}$ lattices \cite{dhar}. R. Qu et.al.
classified equivalence classes in the set of hypergraph states of
three qubits using different entanglement measures \cite{qu}.

One of the measure to quantify three qubit entanglement is
residual tangle originally introduced in \cite{coffman}. Another
measure is called three-tangle which is defined as the square root
of the residual tangle.
 The three-tangle is defined for the pure
three qubit state
$|\psi\rangle=\sum_{i,j,k=0}^{1}a_{ijk}|ijk\rangle$ as
\cite{eltschka}
\begin{eqnarray}
\tau_{3}(|\psi\rangle)= 2 \sqrt{|d_{1}-2d_{2}+4d_{3}|}
\label{tangle}
\end{eqnarray}
where
\begin{eqnarray}
d_{1}&=&a_{000}^{2}a_{111}^{2}+a_{001}^{2}a_{110}^{2}+a_{010}^{2}a_{101}^{2}+a_{100}^{2}a_{011}^{2}\nonumber \\
d_{2}&=&a_{000}a_{111}(a_{011}a_{100}+a_{101}a_{010}+a_{110}a_{001})\nonumber\\
&+&a_{001}a_{110}(a_{010}a_{101}
+a_{011}a_{100})\nonumber\\&+&a_{100}a_{010}a_{011}a_{101}\nonumber\\
d_{3}&=& a_{000}a_{110}a_{101}a_{011}+a_{111}a_{001}a_{010}a_{100}
\label{coeff}
\end{eqnarray}
It can be easily shown that the three-tangle of separable and
biseparable states is zero. The three-tangle for GHZ state is
$2|\lambda\mu|$. When $\lambda=\mu=\frac{1}{\sqrt{2}}$,
$\tau_{3}(|GHZ\rangle)=1$. While the three-tangle for W state is
zero.

The aim of this work is to classify pure three qubit entanglement,
mainly, separable, W class and GHZ class of states by reducing the
three qubit pure states into two qubit mixed states. By analyzing
the quantum correlation of the two qubit mixed state, we can make
definite conclusion on the classification of pure three qubit
states.

The paper is organised as follows: In section-II, we introduced a
special type of two qubit mixed state and studied the quantum
correlation, measured by geometric discord, of the proposed
special two qubit mixed state. In section-III, We study the
purification of the introduced state in section-II and found the
condition of the possibility of the purification. In section-IV,
we calculate the three-tangle of the purified state and then
classify the three qubit purified state. We conclude with a
summary of our result in section-V.

\section{II. Geometric Discord of Special Type of Two-qubit Mixed States}
In this section we have proposed a special type of two qubit
bipartite mixed states $\sigma_{AB}$. The form of $\sigma_{AB}$ is
given by
\begin{eqnarray}
\sigma_{AB}=
p|\chi_{1}\rangle\langle\chi_{1}|\otimes\rho_{1}+(1-p)|\chi_{2}\rangle\langle\chi_{2}|\otimes\rho_{2}
\label{special}
\end{eqnarray}
where $|\chi_{1}\rangle=\alpha|0\rangle+\beta|1\rangle$,
$|\chi_{2}\rangle=\beta|0\rangle+(-1)^{n}\alpha|1\rangle$,
$\rho_{1}=\frac{1}{2}[I_{2}+\vec{r}.\vec{\sigma}]$,
$\rho_{2}=\frac{1}{2}[I_{2}+\vec{s}.\vec{\sigma}]$, $I_{2}$
represent identity matrix of order 2;
$\vec{r}=(r_{x},r_{y},r_{z})$ and
$\vec{s}=(s_{x},s_{y},s_{z})$ represent bloch vectors, $n$ may take odd or even values, and $\alpha^{2}+\beta^{2}=1$.\\

\noindent \textbf{Theorem:} (i) If $n$ is odd, then
$\langle\chi_{1}|\chi_{2}\rangle=0$ and thus $\sigma_{AB}$
represent a zero discord state and hence a classical state.\\
(ii) If $n$ is even, then $\langle\chi_{1}|\chi_{2}\rangle \neq 0$
and thus $\sigma_{AB}$ represent a non-zero discord state and
hence a non-classical state.

\noindent \textbf{Proof:} The state $\sigma_{AB}$ given in
(\ref{special}) can be expanded in terms of pauli-matrices as
\begin{eqnarray}
\sigma_{AB}&=&\frac{1}{4} \lbrace I\otimes I +
\sum_{i=x,y,z}(pr_{i}
+ s_{i}(1-p)) I\otimes\sigma_{i} \nonumber \\
&+& [(2p-1)(\alpha^{2}-\beta^{2})\sigma_{z} +
2\alpha\beta(p+(-1)^{n}(1-p))\sigma_{x}]\nonumber \\
&\otimes& I + p [
(\alpha^{2}-\beta^{2})\sigma_{z}+2\alpha\beta\sigma_{x}]\otimes\overrightarrow{r}.\overrightarrow{\sigma}
\nonumber \\
&-&
(1-p)[(\alpha^{2}-\beta^{2})\sigma_{z}-(-1)^{n}2\alpha\beta\sigma_{x}]\otimes\overrightarrow{s}.\overrightarrow{\sigma}\rbrace
\label{blochsphere}
\end{eqnarray}
In the above Bloch sphere representation
($\overrightarrow{x},\overrightarrow{y}$,T) of $\sigma_{AB}$, the
Bloch vectors $\vec{x},\vec{y}$ is given by
\begin{eqnarray}
& & \vec{x}=(2\alpha\beta(p+(-1)^n(1-p)), 0,
(2p-1)(\alpha^{2}-\beta^{2}),\nonumber \\
& & \vec{y}=(pr_{1}+(1-p)s_{1}, pr_{2}+(1-p)s_{2},\nonumber \\
& & pr_{3}+(1-p)s_{3}) \label{blochvector}
\end{eqnarray}
and the elements of correlation matrix $T$ is given by
\begin{eqnarray}
&& t_{11}=2\alpha\beta[pr_{1}+(-1)^n(1-p)s_{1}],\nonumber \\
& & t_{12}=2\alpha\beta[pr_{2}+(-1)^n(1-p)s_{2}],\nonumber \\
& & t_{13}=2\alpha\beta[pr_{3}+(-1)^n(1-p)s_{3}];\nonumber \\
& & t_{21}=0,t_{22}=0,t_{23}=0;\nonumber \\
& & t_{31}=(\alpha^{2}-\beta^{2})[pr_1-(1-p)s_1],\nonumber \\
& & t_{32}=(\alpha^{2}-\beta^{2})[pr_2-(1-p)s_2],\nonumber \\
& & t_{33}=(\alpha^{2}-\beta^{2})[pr_3-(1-p)s_3]
\label{correlationmatrix}
\end{eqnarray}
The normalised geometric discord of the state $\sigma_{AB}$ is
given by
\begin{eqnarray}
D_{G}(\sigma_{AB})=\frac{1}{2}[\|\vec{x}\|^{2}+ \|\ T
\|^{2}-\lambda_{max}(\vec{x}\vec{x}^{T}+TT^{T})]
\label{discordsigma}
\end{eqnarray}
where $\vec{x}^{T}$ and $T^{T}$ denotes the transposition of
$\vec{x}$ and the correlation matrix $T$ respectively.\\
$\|\vec{x}\|^{2}$ and $\|\ T \|^{2}$ are given by
\begin{eqnarray}
\|\vec{x}\|^{2}&=&4(1-p)^2\langle\chi_1\vert\chi_2\rangle^2+
8\alpha\beta(2p-1)(1-p)\langle\chi_1\vert\chi_2\rangle \nonumber \\
&+& (2p-1)^2 \label{norm1}
\end{eqnarray}
\begin{eqnarray}
\|\ T \|^{2} &=&
\sum\limits_{i=x,y,z}(pr_i-(1-p)s_i)^2+4\langle\chi_1\vert\chi_2\rangle^2(1-p)^2 \times \nonumber \\
&& \sum\limits_{i=x,y,z}s_i^2
+8\langle\chi_1\vert\chi_2\rangle(1-p)\alpha\beta \times \nonumber \\
&& \sum\limits_{i=x,y,z}s_i(pr_i-(1-p)s_i) \label{norm1}
\end{eqnarray}

\noindent The symmetric matrix $\vec{x}\vec{x}^{T}+TT^{T}$ will be
of the form
\begin{eqnarray}
\vec{x}\vec{x}^{T}+TT^{T}=
\begin{bmatrix}
    E & 0 & F \\
    0 & 0 & 0 \\
    F & 0 & G \\
  \end{bmatrix}
\label{symmetric}
\end{eqnarray}
where
\begin{eqnarray}
E&=&4\alpha^2\beta^2[\sum\limits_{i=x,y,z}(pr_i-(1-p)s_i)^2+(2p-1)^2]\nonumber \\
&+&
4\langle\chi_1\vert\chi_2\rangle^2(1-p)^2[\sum\limits_{i=x,y,z}s_i^2+1]\nonumber \\
&+&
8\langle\chi_1\vert\chi_2\rangle(1-p)\alpha\beta[\sum\limits_{i=x,y,z}s_i(pr_i-(1-p)s_i)\nonumber \\
&+&(2p-1)] \label{matrixcoefficient1}\\
F&=&2\alpha\beta(\alpha^2-\beta^2)[\sum\limits_{i=x,y,z}(pr_i-(1-p)s_i)^2+(2p-1)^2]\nonumber \\
&+&2(\alpha^2-\beta^2)\langle\chi_1\vert\chi_2\rangle(1-p)[\sum\limits_{i=x,y,z}s_i(pr_i-(1-p)s_i)\nonumber \\
&+&(2p-1)] \label{matrixcoefficient2}\\
G&=&(\alpha^2-\beta^2)^2[\sum\limits_{i=x,y,z}(pr_i-(1-p)s_i)^2 \nonumber \\
&+& (2p-1)^2] \label{matrixcoefficient2}
\end{eqnarray}
A simple calculation gives us
\begin{eqnarray}
E+G=\|\vec{x}\|^{2}+ \|\ T \|^{2} \label{calculation}
\end{eqnarray}
The maximum eigenvalue of the matrix $\vec{x}\vec{x}^{T}+TT^{T}$
is given by
\begin{eqnarray}
\lambda_{max}(xx^T+TT^T)&=&\frac{1}{2}[(E+G)+\nonumber
\\ && \sqrt{(E-G)^2+4F^2}] \label{eigenvalue}
\end{eqnarray}
Now we consider the following two cases:\\
Case-I: When $n$ is odd, $\langle\chi_{1}|\chi_{2}\rangle=0$. This
reduces the maximum eigenvalue of the matrix
$\vec{x}\vec{x}^{T}+TT^{T}$ to $\|\vec{x}\|^{2}+ \|\ T \|^{2}$ and
it leads the geometric discord $D_{G}$ to zero. Thus if
$\langle\chi_{1}|\chi_{2}\rangle=0$, geometric discord of
$\sigma_{AB}$ is also equal to zero. Hence the state $\sigma_{AB}$
is considered as a classical state.\\\\
Case-II: When $n$ is even, $\langle\chi_{1}|\chi_{2}\rangle\neq0$
and hence in this case it can be easily shown that geometric
discord of $\sigma_{AB}$ is not equal to zero and thus the state
$\sigma_{AB}$ is a non-classical state.

\section{III. Purification of $\sigma_{AB}$}
Purification is a mathematical procedure which associate a (n+1)-
qubit pure state with a n-qubit mixed state. An additional system
is needed in purification procedure and it is known as reference
system or ancillary system \cite{nielsen}.

\noindent \textbf{Theorem:} Let us consider a state $\sigma_{AB}$
given in (\ref{special}).
There exist a purification of $\sigma_{AB}$ if and only if the density matrices described by $\rho_{1}$ and $\rho_{2}$ are pure.\\

\noindent \textbf{Proof:} The spectral decomposition of the
density matrices $\rho_{1}$ and $\rho_{2}$ is given by
\begin{eqnarray}
\rho_{1}=\lambda_{1}^{(1)}|\psi_{1}\rangle\langle\psi_{1}|+\lambda_{2}^{(1)}|\psi_{2}\rangle\langle\psi_{2}| \nonumber\\
\rho_{2}=\lambda_{1}^{(2)}|\psi_{1}^{'}\rangle\langle\psi_{1}^{'}|+\lambda_{2}^{(2)}|\psi_{2}^{'}\rangle\langle\psi_{2}^{'}|
\label{purification}
\end{eqnarray}
$\lambda_{1}^{(1)}=\frac{(1-\sqrt{r_{x}^2+r_{y}^2+r_{z}^2})}{2},\lambda_{2}^{(1)}=\frac{(1+\sqrt{r_{x}^2+r_{y}^2+r_{z}^2})}{2}$
are the eigenvalues of $\rho_{1}$ and
$|\psi_{1}\rangle=\frac{1}{\sqrt{N}}[\frac{r_{z}-\sqrt{r_{x}^2+r_{y}^2+r_{z}^2}}{r_{x}+i
r_{y}}
\vert0\rangle+\vert1\rangle],|\psi_{2}\rangle=\frac{1}{\sqrt{N_{1}}}[\frac{r_{z}+\sqrt{r_{x}^2+r_{y}^2+r_{z}^2}}{r_{x}+i
r_{y}} \vert0\rangle+\vert1\rangle]$ are corresponding orthonormal
eigenvectors.
$\lambda_{1}^{(2)}=\frac{(1+\sqrt{s_{x}^2+s_{y}^2+s_{z}^2})}{2},\lambda_{2}^{(2)}=\frac{(1-\sqrt{s_{x}^2+s_{y}^2+s_{z}^2})}{2}$
are the eigenvalues of $\rho_{2}$ and
$\vert\psi_{1}^{'}\rangle=\frac{1}{\sqrt{N_{1}^{'}}}[\frac{s_{z}+\sqrt{s_{x}^2+s_{y}^2+s_{z}^2}}{s_{x}+i
s_{y}}
\vert0\rangle+\vert1\rangle],\vert\psi_{2}^{'}\rangle=\frac{1}{\sqrt{N^{'}}}[\frac{s_{z}-\sqrt{s_{x}^2+s_{y}^2+s_{z}^2}}{s_{x}+i
s_{y}} \vert0\rangle+\vert1\rangle]$ are corresponding orthonormal
eigenvectors. The normalization constants are given by
$N=1+\frac{(r_{z}-\sqrt{r_{x}^2+r_{y}^2+r_{z}^2})^2}{r_{x}^2+r_{y}^2},
N_{1}=1+\frac{(r_{z}+\sqrt{r_{x}^2+r_{y}^2+r_{z}^2})^2}{r_{x}^2+r_{y}^2},N_{1}^{'}=1+\frac{(s_{z}+\sqrt{s_{x}^2+s_{y}^2+s_{z}^2})^2}{s_{x}^2+s_{y}^2},
N^{'}=1+\frac{(s_{z}-\sqrt{s_{x}^2+s_{y}^2+s_{z}^2})^2}{s_{x}^2+s_{y}^2}$.\\\\
Let a purification of $\sigma_{AB}$ be
\begin{eqnarray}
|\xi\rangle_{ABC}&=&\sqrt{p}|\chi_{1}\rangle_{A}\otimes(\sqrt{\lambda_{1}^{(1)}}|\psi_{1}\rangle_{B}\otimes|0\rangle_{C}
\nonumber \\
&+&\sqrt{\lambda_{2}^{(1)}}|\psi_{2}\rangle_{B}\otimes|1\rangle_{C})+
\sqrt{1-p}|\chi_{2}\rangle_{A}\nonumber \\
&\otimes&(\sqrt{\lambda_{1}^{(2)}}|\psi_{1}^{'}\rangle_{B}\otimes|0\rangle_{C}
+\sqrt{\lambda_{2}^{(2)}}|\psi_{2}^{'}\rangle_{B}\nonumber \\
&\otimes&|1\rangle_{C}) \label{purification}
\end{eqnarray}
where the qubit $C$ represent the ancilla.\\

\noindent If we trace out the ancilla qubit then the reduced density matrix of
the two qubits is given by
\begin{eqnarray}
\varrho_{AB}&=&Tr_{C}(|\xi\rangle_{ABC}\langle\xi|)\nonumber\\
&=&p\vert\chi_{1}\rangle_{A}\langle\chi_{1}\vert\otimes(\lambda_{1}^{(1)}
\vert\psi_{1}\rangle_{B}\langle\psi_{1}\vert\nonumber \\
&+&\lambda_{2}^{(1)}\vert\psi_{2}\rangle_{B}\langle\psi_{2}\vert)
+(1-p)\vert\chi_{2}\rangle_{A}\langle\chi_{2}\vert\nonumber \\
&\otimes&(\lambda_{1}^{(2)}
\vert\psi_{1}^{'}\rangle_{B}\langle\psi_{1}^{'}\vert+\lambda_{2}^{(2)}\vert\psi_{2}^{'}\rangle_{B}\langle\psi_{2}^{'}\vert)
\nonumber \\
&+&\sqrt{p(1-p)}\vert\chi_{1}\rangle_{A}\langle\chi_{2}\vert\otimes(\sqrt{\lambda_{1}^{(1)}\lambda_{1}^{(2)}}
\vert\psi_{1}\rangle_{B}\langle\psi_{1}^{'}\vert\nonumber \\
&+&\sqrt{\lambda_{2}^{(1)}\lambda_{2}^{(2)}}\vert\psi_{2}\rangle_{B}\langle\psi_{2}^{'}\vert)
\nonumber \\
&+&\sqrt{p(1-p)}\vert\chi_{2}\rangle_{A}\langle\chi_{1}\vert\otimes(\sqrt{\lambda_{1}^{(1)}\lambda_{1}^{(2)}}
\vert\psi_{1}^{'}\rangle_{B}\langle\psi_{1}\vert
\nonumber \\
&+&\sqrt{\lambda_{2}^{(1)}\lambda_{2}^{(2)}}\vert\psi_{2}^{'}\rangle_{B}\langle\psi_{2}\vert)
\label{reduceddensity}
\end{eqnarray}
It is clear from equation (\ref{reduceddensity}) that
$|\xi\rangle_{ABC}$ is a purification of $\sigma_{AB}$ iff
$\lambda_{1}^{(1)}=0$ and $\lambda_{2}^{(2)}=0$, i.e. the density
matrices $\rho_{1}$ and $\rho_{2}$ represent pure states.

\section{IV. A Classification of Three Qubit States}
\noindent Let the three qubit states $|\xi\rangle_{ABC}$ is a
purification of $\varsigma_{AB}$
\begin{eqnarray}
\varsigma_{AB}&=& p|\chi_{1}\rangle\langle\chi_{1}|\otimes\rho_{1}
\nonumber\\ &+&
(1-p)|\chi_{2}\rangle\langle\chi_{2}|\otimes\rho_{2},~~0<p<1
\label{special1}
\end{eqnarray}
where $|\chi_{1}\rangle=\alpha|0\rangle+\beta|1\rangle$,
$|\chi_{2}\rangle=\beta|0\rangle+(-1)^{n}\alpha|1\rangle$,
$\rho_{1}=\frac{1}{2}[I_{2}+\vec{r}.\vec{\sigma}]$,
$\rho_{2}=\frac{1}{2}[I_{2}+\vec{s}.\vec{\sigma}]$, $I_{2}$
represent identity matrix of order 2;
$\vec{r}=(r_{x},r_{y},r_{z})$ and $\vec{s}=(s_{x},s_{y},s_{z})$
represent bloch vectors satisfying
$r_{x}^{2}+r_{y}^{2}+r_{z}^{2}=1$ and
$s_{x}^{2}+s_{y}^{2}+s_{z}^{2}=1$.\\
The three qubit states $|\xi\rangle_{ABC}$ can be expressed in
$\{|0\rangle,|1\rangle\}$ basis as
\begin{eqnarray}
|\xi\rangle_{ABC}&=&a\vert001\rangle+b\vert011\rangle+c\vert101\rangle+d\vert111\rangle+e\vert000\rangle \nonumber \\
&+&f\vert010\rangle+g\vert100\rangle+h\vert110\rangle
\label{purificationnew}
\end{eqnarray}
where
\begin{eqnarray}
&& a=\frac{\alpha\sqrt{p}(r_{z}+1)(r_{x}-i
r_{y})}{\sqrt{(r_{x}^{2}+r_{y}^{2})(r_{x}^{2}+r_{y}^{2}+(r_{z}+1)^{2})}},\nonumber
\\ && b=\frac{a(r_{x}+i r_{y})}{(r_{z}+1)},~~~~~ c=\frac{a\beta}{\alpha} \nonumber
\\ && d=\frac{a\beta}{\alpha}.\frac{(r_{x}+i
r_{y})}{(r_{z}+1)}\nonumber
\\ && e=\frac{\beta\sqrt{(1-p)}(s_{z}+1)(s_{x}-i
s_{y})}{\sqrt{(s_{x}^{2}+s_{y}^{2})(s_{x}^{2}+s_{y}^{2}+(s_{z}+1)^{2})}}\nonumber
\\ && f=\frac{e(s_{x}+i
s_{y})}{(s_{z}+1)},~~~~~ g=(-1)^{n}.\frac{e\alpha}{\beta}\nonumber
\\ && h=(-1)^{n}.\frac{e\alpha}{\beta}\frac{(s_{x}+ i
s_{y})}{(s_{z}+1)} \label{coefficientthreequbit}
\end{eqnarray}
The three qubit entanglement contained in the state
$|\xi\rangle_{ABC}$ is measured by 3-tangle and it is given by
\begin{eqnarray}
\tau_3(\vert \xi \rangle_{ABC}) = 2 \sqrt{| k_{1} - 2k_{2} +
4k_{3}|} \label{threetangle}
\end{eqnarray}
where
\begin{eqnarray}
k_{1} &=& e^{2}d^{2} + a^{2} h^{2} + f^{2}c^{2} +
b^{2}g^{2}\nonumber
\\ k_{2} &=&
ed(ah + fc + bg) + ah(fc + bg) + fbgc \nonumber
\\ k_{3}&=& bceh + adfg \label{threetanglemeasure}
\end{eqnarray}
A simple calculation and after some simplification, the 3-tangle
gives
\begin{eqnarray}
\tau_3(\vert \xi \rangle_{ABC}) =
\frac{2|ae(\alpha\langle\chi_{1}|\chi_{2}\rangle-\beta)|}{\alpha\beta^{2}}\Delta\label{threetanglemeasure1}
\end{eqnarray}
where $\Delta = \sqrt{(\frac{s_x}{1 + s_z} - \frac{r_x}{1 + r_z})^2 + (\frac{s_y}{1 + s_z} - \frac{r_y}{1 + r_z})^2}.$\\

\textbf{Case-I:} If $ae=0$ holds then $\tau_3(\vert \xi
\rangle_{ABC})=0$ for any non-zero values of
$\langle\chi_{1}|\chi_{2}\rangle$. In this case the state $|\xi
\rangle_{ABC}$ is a biseparable state.\\

\textbf{Case-II:} If $\Delta=0\Rightarrow
\frac{s_{x}}{r_{x}}=\frac{s_{y}}{r_{y}}=\frac{1+s_{z}}{1+r_{z}}$
and $ae=0$ then the three tangle $\tau_3(\vert \xi
\rangle_{ABC})=0$ for any non-zero values of
$\langle\chi_{1}|\chi_{2}\rangle$. Hence the state $|\xi
\rangle_{ABC}$ is a biseparable state.\\

\textbf{Case-III:} If $\Delta=0$ and $ae\neq 0$ then the three
tangle $\tau_3(\vert \xi \rangle_{ABC})=0$ for any non-zero values
of $\langle\chi_{1}|\chi_{2}\rangle$. Thus the state $|\xi
\rangle_{ABC}$ is a W-type state.\\

\textbf{Case-IV:} If $\Delta\neq 0$, $ae\neq 0$ and $\langle
\chi_1 \vert \chi_2 \rangle \neq
\frac{\beta}{\alpha}(\textrm{i.e.}~ \alpha\neq\frac{1}{\sqrt{2}})$
then the three qubit state
$|\xi \rangle_{ABC}$ represent GHZ-type state and the amount of entanglement is given by (\ref{threetanglemeasure1}).\\

\section{V. Conclusion}
\noindent In this work, we have investigated the different types
of entanglement classification of purified tripartite state
$|\xi\rangle_{ABC}$ of the two qubit mixed state $\varsigma_{AB}$
based on geometric discord of $\varsigma_{AB}$. We find that the
three qubit state of the type biseparable, W-type and GHZ-type
state can be generated from non-classical two qubit mixed state.
We can also quantify the amount of entanglement in the generated
GHZ-type state.

\end{document}